# Dev-for-Operations and Multi-sided Platform for Next Generation Platform as a Service

Bela Berde[1], Steven Van Rossem[2], Aurora Ramos[3], Matteo Orrù[4], Anas Shatnawi[4]
[1]Nokia Bell-Labs France, [2]Ghent University-imec, [3]ATOS Spain, [4]University of Milano-Bicocca

*Abstract*— This paper presents two new challenges for the Telco ecosystem transformation in the era of cloud-native microservice-based architectures. (1) Development-for-Operations (Dev-for-Operations) impacts not only the overall workflow for deploying a Platform as a Service (PaaS) in an open foundry environment, but also the Telco business as well as operational models to achieve an economy of scope and an economy of scale. (2) For that purpose, we construct an integrative platform business model in the form of a Multi-Sided Platform (MSP) for building Telco PaaSes. The proposed MSP based architecture enables a multi-organizational ecosystem with increased automation possibilities for Telco-grade service creation and operation. The paper describes how the Dev-for-Operations and MSP lift constraints and offers an effective way for next-generation PaaS building, while mutually reinforcing each other in the Next Generation Platform as a Service (NGPaaS) framework.

## I. Introduction

The transformation of Telco infrastructures into cloud-native and microservice-based architectures calls to redesign compute, storage, and network components beyond the current virtualization technologies [1]. Microservices extend cloud-native concepts. Being cloud-native, microservices are interchangeable, replaceable, and composable in a more than one supplier environment. Accordingly, microservice-based Telco applications are no longer seen as stand-alone software but as aggregates from third-party microservices, which are not simple add-ons to the core application. Building and deploying Telco applications in an infrastructural agnostic way has therefore the following main features:

1. *composability*: with the so-called "telco-grade" quality, Telco applications are enabled to combine all sort of third-party microservice-architected software components for creating new, versatile, and powerful cloud objects, with no silos between connectivity, storage, and computing units.
2. *PaaS*: when deployed, such an application is called an application Platform-as-a-Service (PaaS), ideal to remove all the repetitive tasks encountered on top of the Infrastructure-as-a-Service (IaaS) layer, as IaaS resources automatically scale. An individual PaaS instance may cover multiple IaaS segments, where each individual part of the PaaS - running on a given IaaS segment - is reprogrammable (Figure 1).
3. *reusability*: combining a rich menu of telco-grade applications as PaaS instances raises the question of multi-vendor microservice and, more generally, component reusability for cloud-computing-assisted large-scale distributed production. Reusable components free the development, deployment, and operation of PaaSes from reimplementing the same microservice systems over and over again.
4. *automation*: while pipelines vary from organization to organization, as extra value-added services, PaaS instances require a complete set of automation technologies for component integration, building, deployment, and technical operations.
5. *platform*: automated building and operation of PaaS instances call for an industry platform, which changes the relation between microservice developers, service providers, and end-users of those services.

The next-generation Telco PaaS, including features 1.-5., is, by definition, cloud-native, componentized, and are operated in a multi-organizational environment in an automated way.

Moreover, combining microservices, coming from various sources and firms, requires a building and deployment strategy, i.e., a PaaS software foundry approach. The approach demands a methodology to simplify and accelerate the PaaS production, especially, within an increasingly complex and dynamic software and hardware production context [2].

The methodology implies a new definition of what is a reusable software component as fundamental building block as well as the concept of ecosystem, tooling chains, and business model transformation:

1. *Reusable Functional block (RFB)*: the microservice concept needs to be transformed into Reusable Functional Blocks (RFB). Simply stated, an RFB is defined as a microservice augmented with the declarative statement of some functional, performance, and execution environment metadata for describing the link between functionality and infrastructure [4]. The declarative statement is called blueprint. It contributes to automating the building, operation and orchestration strategy of the RFB with some metadata such as the description of the desired deployment target. Further, the RFB can be seen as a broader reusable component concept, by which not only microservices but also bunches of microservices and even services are concerned: it is not only a generic unit of composition but also a recursive element. This raises the issue of the visibility on RFB internal components, for example.
2. *Development-for-Operations (Dev-for-Operations)*: tools and techniques applied in a whole ecosystem for handling RFBs.
3. *Multi-sided Platform (MSP)*: platformization implies a new ecosystem for organizing collaboration and control on

software, hardware, and services. It also demands the redesign of the business models in the form of a multi-sided platform (MSP) [3]. Without taking ownership of the services whose exchanges it facilitates in inter-organizational interactions, an MSP is both an industry platform and an intermediary. The utilization differential of the MSP appears in both a potential economy of scope and in the opportunity for economy of scale.

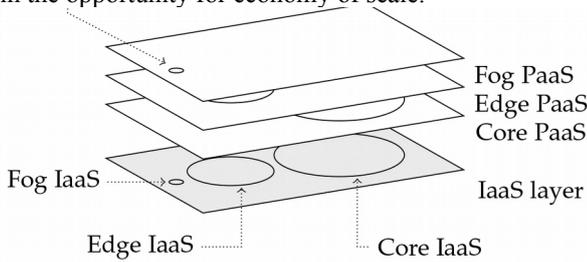

Figure 1: PaaS instances for a set of Fog, Edge, and Core IaaS instances.

## II. CURRENT PaaS MODEL

Current PaaS solutions, as an independent service model for the provision of a complete platform, i.e. hardware and software as service bunches, propose to develop software applications and to integrate them with infrastructure for deployment. The platform provides the platform users with all functionalities which are needed during the lifecycle of an application, from development and testing to deployment and operations. PaaS solutions propose extra value-added services such as support, monitoring, lifecycle management, quality assurance, and certification. Most PaaS solutions have the drawback of applications being tightly locked to the PaaS provider development and runtime environment, namely an integrated development and deployment environment that normally supports the use of multiple but selected programming languages and offers crafted tools for implementing, testing, and operating. Therefore, the application relies on the offered tools by the PaaS provider, who can also propose a marketplace with catalog, rating, usage tracker, recommendation, pricing, and billing for platform usage. Activities on current PaaS solutions are not planned to be multi-organizational.

## III. TELCO PaaS AND DEV-FOR-OPERATIONS

A next-generation Telco PaaS instance, on the other hand, can be considered as an RFB by extension, allowing on-demand PaaS setup in each PaaS segment (Figure 1). Individual PaaS instances are further reprogrammable and customizable to meet the requirement of services to be deployed. Using many different RFBs, PaaS instances are tailored to the needs of a wide range of use cases with telco-grade 5G characteristics.

To develop and to operate PaaS instances using traditional DevOps processes is not enough for an industry platform. While DevOps tools and techniques became customary as applied to an entire organization in the IT industry, multi-organizational software development and operation is a broader concept than DevOps. In siloed organizational environments, DevOps allows essentially changing the roles into communities of practice. DevOps ties streamlined release pipelines for automated connecting of different software development activities performed by several teams.

However, multi-organizational communities of practice tend to complexify. In order to exert control over the multi-organizational software/hardware production system at any stage, Dev-for-Operations practices allow inter alia, multi-organizational shared mutual understanding, shared work goals, shortened feedback cycles, and a collaborative environment as opposed to a competitive one. Dev-for-Operations is enforced by specific tools for continuous monitoring of the RFB lifecycle in a whole ecosystem within the industry platform.

The Dev-for-Operations challenge is to surmount the limitations imposed by and combine the advantages of the integrated development environment individualized for software/hardware vendors, the service building environment of service providers, and the production environment of the platform provider, while enhancing the extent by which two implementations or two components from different manufacturers can co-exist and work together (Figure 2). The Dev-for-Operations enabled industry platform naturally leads to a new business model and to the MSP adoption.

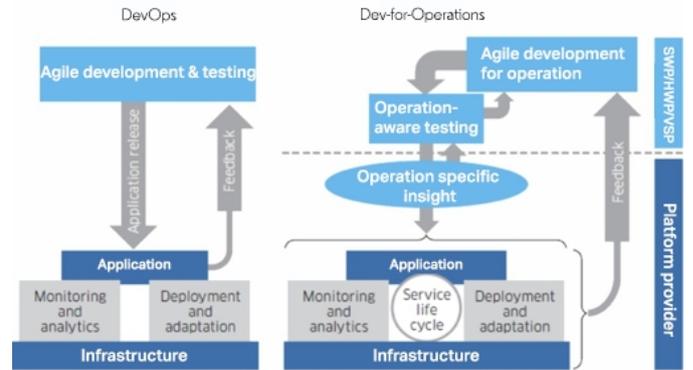

Figure 2: Difference between DevOps and Dev-for-Operations [5].

## IV. TELCO PaaS AND MSP

The RFB-based system is essentially a system that offers components that can easily be used by, or integrated into other RFBs. At the same time, the system itself will often consist of components that originate from elsewhere. Flexibility in interoperability, composability, and extensibility of components is therefore a key feature. A new RFB and service production and operation model is needed that, from a business engagement perspective, fundamentally transforms the interaction with component providers and end-users. This makes the industry platform provider role central.

To successfully build for flexibility, the platform provider offers tools and processes to redesign the industry platform as a multi-sided marketplace [6]. The common tools and processes provided by the MSP include individualized workspaces for participants, building, testing, and delivery tools and repositories for RFBs, and operation monitoring and analytics functionalities, where participants, RFBs, and processes have a corresponding OSS/BSS counterpart (Figure 3).

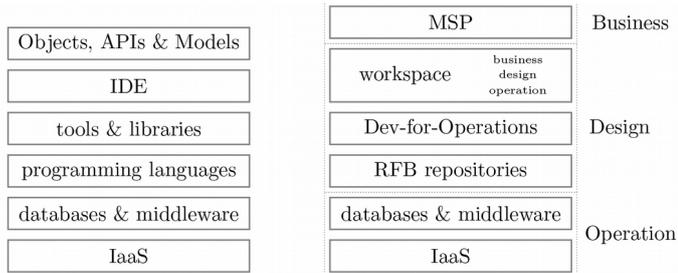

Figure 3: Current PaaS stack versus NGPaaS stack.

The MSP allows a multi-directional flow of value between different participants:
- from component point of view: the software provider (SWP) side and the hardware provider (HWP) side. The SWP side includes Telco manufacturers, software houses, and Free and Open-Source Software (FOSS) developers. The hardware provider (HWP) side is formed by participants, such as, inter alia, open source hardware acceleration providers, hardware manufacturers, public cloud providers, custom hardware acceleration providers, and Artificial Intelligence-as-a-Service (AIaaS) providers running hardware.
- from service point of view: the services composed and deployed through PaaS instances, i.e., RFBs, by the vertical service provider (VSP) side to meet the end-user (EU) needs.

The MSP is, therefore, located between a VSP participant and the end-users of that VSP as well as between the platform provider and its own end-user base. The different end-users form the EU side. This is why the MSP is different from the one stop shop Network Function Virtualization (NFV) and the current PaaS solutions marketplaces. The end-users are kept independent from the NFV or PaaS solution marketplaces.

Given the double practicality, namely (1) building component RFBs using software and hardware components, and (2) operating services RFBs to end-users, the MSP has a bi-core architecture base, formed by (1) SWP and HWP and (2) VSP and EU (Error: Reference source not found4).

As coordinator for Dev-for-Operations, the MSP manages different sides, accessibility to RFB categories, individualized workspaces for the different RFB production processes as well as dedicated workspaces for RFB operation. For that reason, it manages Dev-for-Operations pipelines, and further exerts its control, in general, on business interactions.

All the RFBs are built within the MSP. Building is essentially an RFB composition task and is performed by using an individualized workspace formed by business, design, and operation views (Figure 3). The RFB composition includes free and open-source software (FOSS) RFBs or RFBs located in repositories internally available (Figure 3).

In this **build phase**, the RFB production includes RFB testing stages, RFB composition, if any, tested RFB image creation, and, finally, deployment test with specification of the tested execution environment that mimics the final deployment environment, but is completely dedicated to the RFB editor participant. The RFB then put by the RFB editor in an RFB production repository with some accessibility policies for RFB re-use by other participants (Figure 3). Finally, in the **ship phase**, the RFB image is transferred from the production repository of the RFB editor to a ship repository that can be shared by all the participants of the MSP.

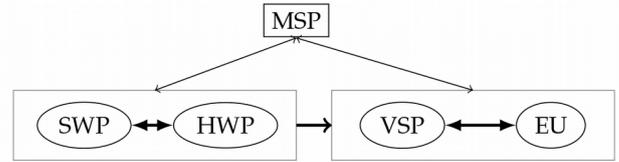

Figure 4: Bi-core architecture of the service provider MSP sides (periphery not shown).

As an example, a service RFB is composed in the workspace of a VSP participant, that allows experimenting for service building, shipping, and running (Figure 3). Once the service is tested and validated, i.e., available in a ship repository, it can be deployed through a PaaS. This is performed, for instance, by a second VSP participant, who transfers the service RFB to the service deployment repository that is common to the platform. The run phase corresponds to deploying and running service RFBs in the appropriate target environment. The PaaS, onboarding the service RFB, can therefore be seen as an enabler for service RFB run operations, with related configuration and binding to IaaS.

V. SERVICE RFB DEPLOYMENT IN THE NEXT-GENERATION PAAS

The different contributions and interactions between the MSP participants, enabled by Dev-for-Operations related functionalities, were previously exemplified in the overall multi-organizational development and operation process. These interactions are explained further using the high-level architectural blocks of the NGPaaS framework (Figure 5). NGPaaS is architected as a cloud-native implementation based upon MSP principles. As shown on Figure 1, different subsets of the IaaS are allocated to several specialized PaaS instances, which is further enhanced with the following functional layers:
- **Business Layer:** The MSP participants register first in this layer. All access, execution rights, license management is regulated and configured accordingly in the layers below for each participant, e.g., by providing specific login credentials coupled to custom monitoring capabilities. At affiliation time, a participant defines its initial role. The role can be transformed on demand into another role affecting the business relationships within the MSP. In this way, in the SWP side, for example, a participant can develop and test its own software components and deploy them later as a VSP participant. Furthermore, BSS functionalities for IaaS, PaaS, and RFB usage and Dev-for-Operations processes are grouped in this layer.
- **Operation Layer**: The operational aspects of deploying PaaS as service components are handled here. This includes automated orchestration of (i) PaaS components to the allocated IaaS and (ii) service components to a supporting PaaS. The layer also contains OSS functionalities for the Dev-for-Operations, PaaS, and IaaS layers.
- **Dev-for-Operations Layer**: This is part of the RFB build, ship, and run environment as explained earlier. This layer implements the different interfaces needed for Telco-grade development cycles, such as customizable monitoring and multi-organizational integrations.

(and isolated), for instance, HWP participant provided resources before deployment. Southbound, monitored data can be gathered to assess the correct functionality of the component.

5. Once validated by the Dev-for-Operations layer, the SWP participant's component can be onboarded by the platform in a way it is available for production usage in the future, e.g., by including the component in the appropriate marketplace catalog.

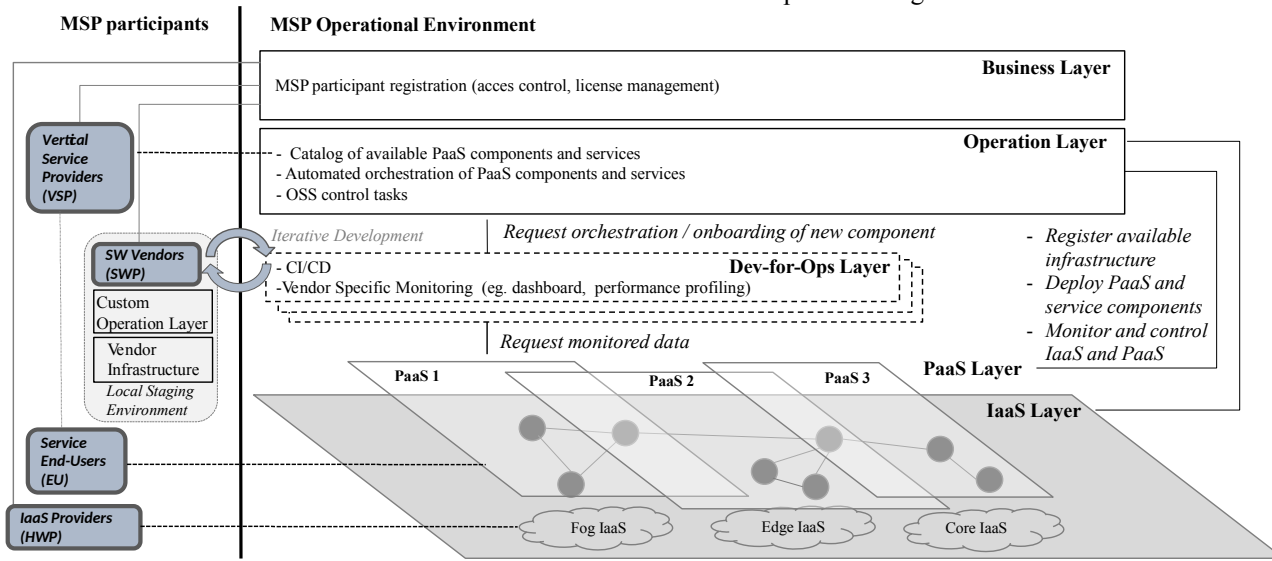

Figure 5: High-level architectural blocks which support the MSP and Dev-for-Operations based interactions between involved MSP participants.

Using these functional layers, we further explain how the interactions between the MSP participants take place in this architecture (Figure 6).

### A. Dev-for-Operations Interactions

Softwarized PaaS or service components are provided by the SWP side and onboarded through the Dev-for-Operations Layer. An iterative development cycle is enabled using following principles (and illustrated in Figure 6):

1. The SWP participant creates a local Design environment, a workspace, to build its components.
2. A local instance of the Operation Layer is deployed by the SWP participant, where the software component can be tested using similar orchestration and control mechanisms as used in production. This is a customized version, allowing the SWP to deploy a selected PaaS instance on its own managed infrastructure layer.
3. Once validated locally, the SWP participant transfers the software component through the Dev-for-Operations layer for deployment. Northbound, the Dev-for-Operations layer can request the Operation Layer to deploy the new component.
4. This triggers a Continuous Integration/Continuous Development (CI/CD) process, where the component is validated through several integration tests on appropriate

Above procedure describes how the SWP participant can validate its components with the HWP participants and platform provider before they are accepted by the MSP. Once the platform onboards the new or updated software component, it can be deployed in a production PaaS environment or offered to a VSP via the MSP service catalog. In this way, the MSP environment plentily plays its intermediary role and behaves as an interface to support the needed interaction chains.

### VI. MONITORING IN DEV-FOR-OPERATIONS

In the Dev-for-Operations model the *dev* side cannot interfere directly with the *ops* environment for services running on top of a PaaS. The SWP/HWP/VSP are even unaware where their services are orchestrated to. Monitoring is, however, a crucial step already in any DevOps cycle so the *dev* side can get the necessary inputs from monitored data during operation, relevant to analyze and debug RFBs.

In addition, different MSP sides have different monitoring demands: VSPs want to provide a monitoring service to their EU, who wants to be monitored on its own, companies might need to provide an evidence to their prospects of how reliable their services are, and SWPs want to monitor a newly deployed version of their software. All can co-exist in the same PaaS. These are just few examples of complexity in a heterogeneous PaaS environment. These scenarios call for reliable, flexible and highly customizable monitoring functionalities which also needs to be scalable.

Therefore, the Dev-for-Operations layer allows RFB producers for customizable configuration of limited access into the operational environment of PaaS instances, where the services are deployed.

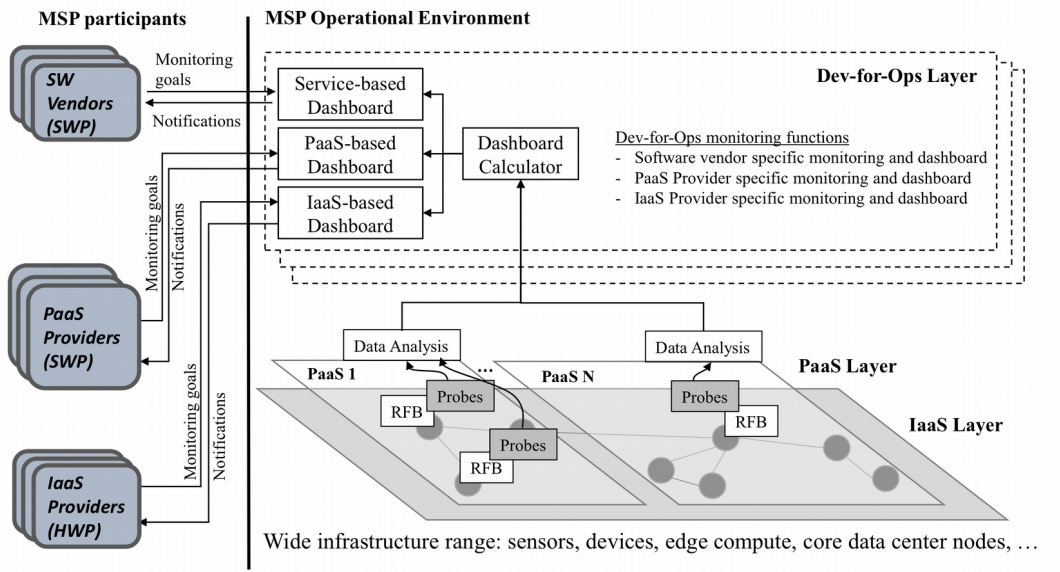

Figure 6: Dev-for-Operations monitoring and dashboard, specific for different vendors and providers.

In this way, the SWP/HWP/VSP can access monitoring data, isolated from the monitoring data related to other RFBs. In order to share the monitored data from the operational environment of the platform to the development side - and especially get the metrics in real-time related to the operational health status - a Dev-for-Operations-related interface is foreseen by means of a Dashboard for each SWP/HWP/VSP (Figure 6). Therefore, different monitoring purposes might co-exist in the platform for an RFB that further represent the complexity of the heterogeneous cloud environment [8].

The capability to gather monitored data is the basis for profiling RFB performance. The use-cases for monitoring in Dev-for-Operations is extended with data analytics in order to characterize the expected performance of an RFB under a certain workload and varying IaaS resources. The Dev-for-Operations layer could then assist the SWP participant to provide recommendations for resource allocation. The modeled performance of an RFB will assist greatly in RFB resource estimation and capacity planning, including a better prediction of the scaled in/out service performance [7].

## VII. CONCLUSIONS

Following the deep infrastructure transformation, the key difference between cloud-native and microservice- or RFB-based architectures increases the need for fully automated PaaS building. Automation becomes the driver for innovative cross-organization business models, which will impact the entire ecosystem and business logic. Comparing the one stop shop NFV marketplace model and the MSP model shows that NFV marketplaces are only intermediary.

The Telco-grade PaaS foundry implemented in NGPaaS with MSP and Dev-for-Operations practices can create communities within the ecosystem that unlock radically new service production and automation opportunities.

ACKNOWLEDGMENT

The work is done in the NGPaaS, Next Generation Platform as a Service project, founded by the European Commission, H2020-ICT-2016-2, ICT-8-2016.